\documentclass[twocolumn,preprintnumbers,amsmath,amssymb]{revtex4}

\usepackage{graphicx}% Include figure files
\usepackage{dcolumn}% Align table columns on decimal point
\usepackage{bm}% bold math

\begin{document}

%\preprint{To be submitted to Rev. Sci. Instrum.}

\title{Construction of a Versatile Ultra-Low Temperature Scanning
Tunneling Microscope}

\author{H. Kambara}
\author{T. Matsui}
\author{Y. Niimi}
\author{Hiroshi Fukuyama}
%\email{hiroshi@phys.s.u-tokyo.ac.jp}
\affiliation{Department of Physics, Graduate School of Science,
The University of Tokyo,
7-3-1 Hongo, Bunkyo-ku, Tokyo 113-0033, Japan}


\date{\today}% It is always \today, today,

\begin{abstract}
We constructed a dilution-refrigerator (DR) based ultra-low temperature
scanning tunneling microscope (ULT-STM) which works at temperatures down
to 30 mK, in magnetic fields up to 6 T and in ultrahigh vacuum (UHV).
Besides these extreme operation conditions, this STM has several
unique features not available in other DR based ULT-STMs.
One can load STM tips as well as samples with clean surfaces
prepared in a UHV environment to an STM head keeping low temperature
and UHV conditions.
After then, the system can be cooled back to near the base temperature
within 3 hours.
Due to these capabilities, it has a variety of applications
not only for cleavable materials but also for almost all conducting materials.
The present ULT-STM has also an exceptionally high stability in the presence
of magnetic field and even during field sweep.
We describe details of its design, performance and applications
for low temperature physics.
\end{abstract}

%\pacs{71.70.Di, 73.61.Wp, 68.37.Ef, 71.20.Tx}
\maketitle

% body of paper here
%%%%%%%%% Introduction %%%%%%%%%%%%%%%%%%%%%%%%%
\section{INTRODUCTION}
The scanning tunneling spectroscopy (STS) technique by use of scanning
tunneling microscope (STM) has become important for studies of
low temperature physics in recent years.
The STS is a powerful method to investigate local electronic density of
states (LDOS) of material surfaces with high spatial ($\le0.1$ nm)
and energy resolutions ($<1$ meV) at cryogenic temperature below 10 K.
Potentially, there are many applications in that temperature range,
for example, unconventional superconductors with low transition
temperatures ($T_{\rm c}$), quantum nanostructures on semiconductor
surfaces, $etc$.
In principle, we can expect better performance of STS at lower
temperatures because of smaller thermal broadenings of tunnel spectra
and thermal expansion coefficients of materials.
So far, several STMs that work at sub-Kelvin temperatures have been
constructed; for example, $^3$He refrigerator based STMs
($T\ge0.3$ K) \cite{pan,geneve,hamburg} and dilution refrigerator (DR)
based STMs ($T\le0.1$ K) \cite{hess,tsukuba,delft,cnrs,madrid,hanaguri}.
However, due to technical difficulties, none of the existing DR-based
STMs has the {\it full} ultrahigh vacuum (UHV) compatibility
with which one can prepare clean sample surfaces {\it in situ}.
In other words, samples are limited only to cleavable materials
or those with chemically inert surfaces for so far constructed STMs
based on DR.

In this article, we describe details of design, construction and test
results of a versatile ultra-low temperature STM (ULT-STM) which we
have constructed recently \cite{ourSTM1,ourSTM2}.
It works at triple extreme conditions, $i.e.$, at very low temperatures
down to 30 mK, in high magnetic fields up to 6 T and in UHV.
It has also the following unique features: one can (1) prepare clean
sample surfaces in a UHV chamber, (2) load the samples and tips
to an STM head keeping low temperature and UHV conditions,
and then (3) cool them back to near the base temperature within 3 hours.
Therefore, the ULT-STM has a variety of applications without being restricted
only to cleavable materials such as layered ones bound by the van der Waals
force. It has been successfully used, for example, in STS measurements
of Landau quantization in two-dimensional electron systems at surfaces
of semiconductors \cite{niimiinas} and
graphites \cite{matsuiprl, niimipe, niimiprl} and of surface electronic
states in a spin-triplet superconductor Sr$_2$RuO$_4$ \cite{kambara}.

%%%%% Experimental Techniques %%%%%%%%%%%%%%%%%%%%%%%
\section{DESIGN AND CONSTRUCTION}
\subsection{Cryostat with bottom-loading mechanism}

\begin{figure}[h]
\includegraphics[width=0.9\linewidth]{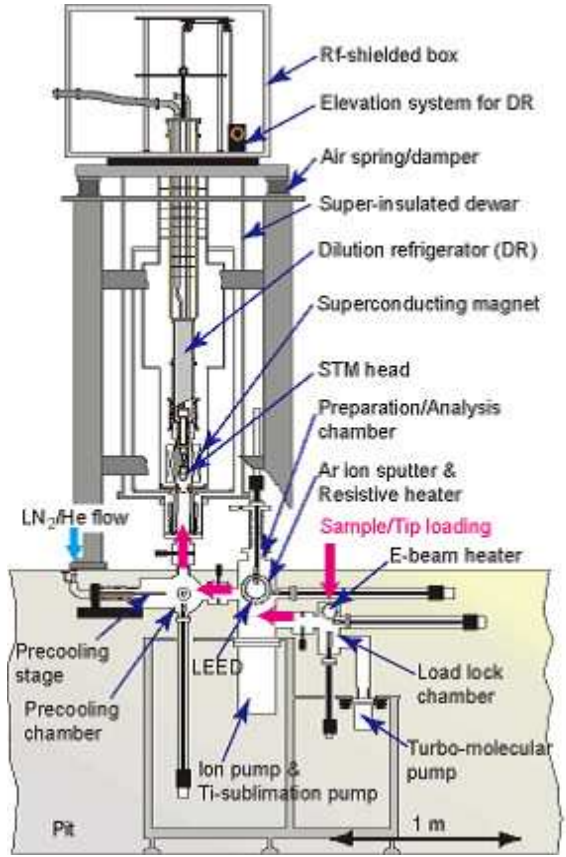}
\caption{\label{overview}
Overview of the ULT-STM. The thick arrows denote the loading path of
samples and tips.}
\end{figure}

For sample/tip exchange, we adopted the bottom-loading mechanism,
instead of the ordinary top-loading one because of the following technical
reasons.
(1) One can shorten the access length for a sample/tip transfer rod.
(2) A specially designed DR with a large-bore central access for
top-loading is not necessary.
UHV chambers for sample/tip preparations and evaluations are located
on floor of an experimental pit of 1.5 m deep.
The cryostat is hung from an anti-vibration table and connected
with the UHV chambers at the bottom (Fig.~\ref{overview}).
The central part of the cryostat is a commercial DR \cite{oxford}
with a super-insulated dewar.
The volume of liquid helium reservoir (74 liters) is large enough to
maintain the base temperature for almost 3 days without refilling
liquid helium.
A 6 T superconducting magnet with a 90 mm bore diameter is immersed in
the liquid-helium bath and surrounds a UHV can (see Fig.~2).
All electronic equipments such as a current preamplifier \cite{preamp}
and a controller \cite{stmcu} for STM measurement are located in an
rf-shielded box/room in order to eliminate electrical noises from outside.

\begin{figure}[h]
\includegraphics[width=0.9\linewidth]{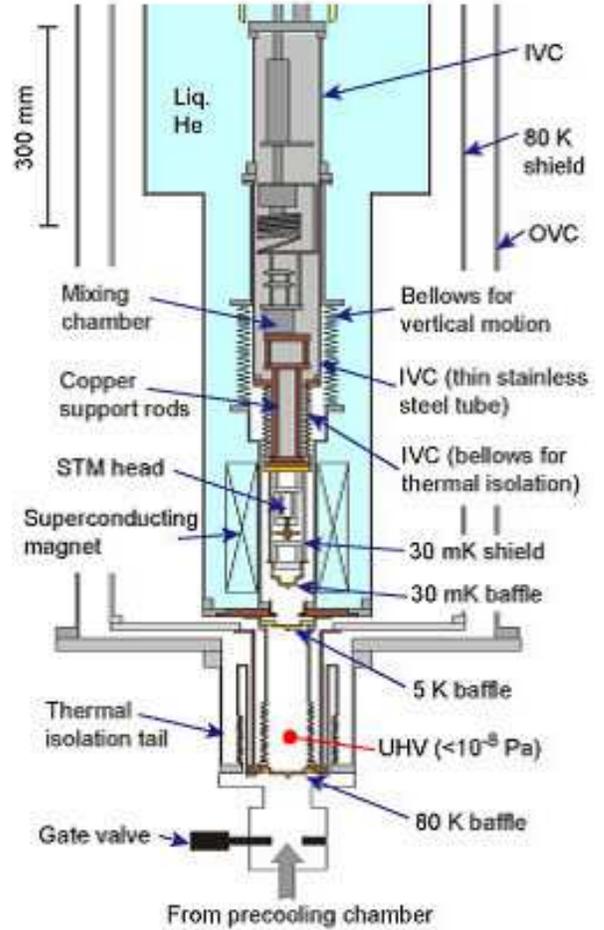}
\caption{\label{cryostat}
Central part of the cryostat. The STM head is located in the UHV space
separated from the vacuum space for DR (light grey region).}
\end{figure}

\subsection{STM head}
\begin{figure}[h]
\includegraphics[width=1.0\linewidth]{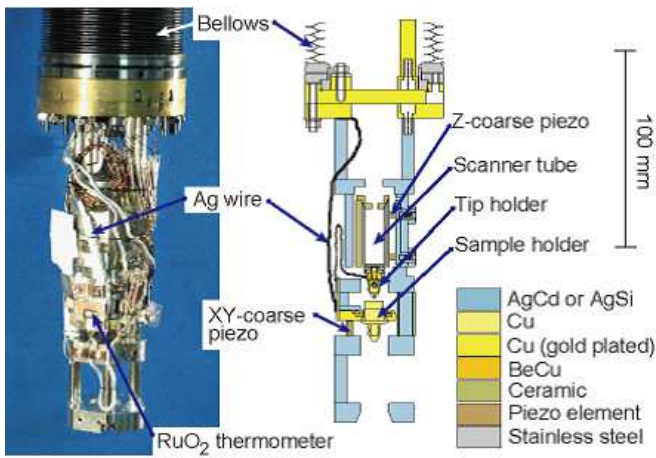}
\caption{\label{stmhead}
Photo (left) and cross sectional drawing (right) of the STM head.}
\end{figure}

A photo and a cross-sectional view of the STM head \cite{unisokustm}
are shown in Fig.~\ref{stmhead}.
The main body of the head is made of hard silver
(Ag($>$ 99 at.\%)+Cd ($<$ 1 at.\%)) and silicon silver
(Ag$_{0.85}$Si$_{0.15}$) rather than copper or copper-based alloys
to prevent large nuclear-spin heat capacities
in high magnetic fields at very low temperatures.
These silver-based alloys have better mechanical strength compared to
pure metals and proper thermal conductivity \cite{fukuyama}.
On the other hand, a sample stage, sample holder and tip holder
are made of high purity copper for better thermalization
and moderate mechanical strength.
They are thermally linked to the mixing chamber (MC)
of DR through annealed silver foils and wires (see next section).

The coarse approach of tip towards sample surface (Z-coarse approach)
is obtained by the {\it stick and slip} motion with piezo actuators.
Two of the actuators hold an inner ceramic tube,
which contains a single-tube piezo scanner, on three sides of the triangular
cross-section.
The speed of Z-coarse approach is 60 $\mu$m/s with an application
of pulses of $\pm$150 V at 0.5 kHz at temperatures below 4 K.

The tip holder is fixed to one end of the piezo scanner
with an M3 screw.
The maximum XY scan area of the scanner is
1.5$\times$1.5 $\mu$m$^2$ below 4 K, while
4.5$\times$4.5 $\mu$m$^2$ at room temperature.
The sample holder is also screwed into the sample stage with M12.
Samples are glued onto the sample holder with silver epoxy.
For thermometry, a RuO$_2$ chip-resistor is attached to the sidewall of
sample stage.

\subsection{Thermal contacts and isolations}
The UHV space for the STM head is separated from the vacuum
space for DR.
Therefore, the inner vacuum can (IVC) is made of thin-wall stainless
steel tubes (0.5 mm thick) and welded bellows (0.12 mm thick)
to reduce heat conduction from one end at 4 K to another
at the base temperature (see Fig.~\ref{cryostat}).
The STM head is rigidly supported from the bottom of the IVC and
thermally connected to the MC with three support rods
made of copper (8 mm in diameter and 168 mm long).
The STM head is surrounded by a copper radiation shield with a
demountable bottom cap which is thermally anchored to the MC.

The sample and tip are thermally anchored independently to the MC
with additional well-annealed high-purity silver thin-foils and -wires.
Mylar sheets (25 $\mu$m thick) are inserted between the sample
stage/tip holder and the silver thermal links for electrical isolation.

In order to minimize the access length for bottom loading, we made
a specially designed attachment ({\it thermal isolation tail})
which separates the UHV space from the vacuum space for the dewar
inside the outer vacuum can (OVC) (see Fig.~\ref{cryostat}).
The thermal isolation tail consists of three coaxial cylinders to
minimize heat leaks from higher temperature sides.
Several parts of the inner and outer cylinders are made of
thin-wall stainless-steel tubes (0.5 mm thick) and flexible tubes
(0.15 mm thick).
The middle cylinder made of copper is thermally anchored to
the neck of the dewar ($T\approx80$ K).
Between the cylinders, 30--50 layers of aluminized Mylar sheets
(super-insulation) are inserted for radiation shielding.
Two demountable baffles thermally anchored to 5 K and 80 K shield
radiation through the access line from the bottom.

To reduce heat leaks through electrical lead wires,
we used two different kinds of wires in the IVC;
a co-axial cable (inner conductor: superconducting NbTi wire of
0.1 mm in diameter,
outer conductor: CuNi tube of 0.6 mm in diameter)
for tunnel current detection or the bias voltage application
and superconducting NbTi twisted-wires (each 0.10 mm in diameter)
for the high voltage application to the piezo actuators.
In the UHV space, all the wires are
converted to insulated copper twisted-wires which are thermally
anchored to the MC.

\subsection{UHV chambers}
The UHV chamber \cite{unisoku} consists of three parts:
(1) load-lock chamber, (2) preparation/analysis chamber, and
(3) precooling chamber (see Fig.~\ref{overview}).
The actual procedure of sample/tip preparation is as follows.
At first we introduce a new sample or tip into the load-lock chamber and
evacuate it below 10$^{-5}$ Pa by a turbo molecular pump. The tip can be
cleaned by electron beam heating, if necessary.
Then it is transferred to the preparation/analysis chamber with
a magnetic transfer rod.
We can prepare a clean sample surface here by resistive heating and/or
argon ion sputtering in a UHV environment ($\approx10^{-8}$ Pa)
achieved with an ion pump and a titanium sublimation pump
(1600 $\ell/s$).
The surface quality can be analyzed by low-energy electron diffraction
(LEED).
After the surface preparation, the sample/tip is transferred to a
copper platform in the precooling chamber.
Here it is cooled by liquid nitrogen or liquid helium flow
down to 100 K or 7 K, respectively.
The sample cleavage can be done at various temperatures between 7 and 300 K
by pushing a small post glued on top of the sample with a blade
in this chamber.
Finally, the precooled sample/tip is transferred to the STM head.

\subsection{Vibration isolation}
The whole cryostat is suspended from the anti-vibration table with air springs,
dampers (vertical direction) and rubber stacks (horizontal direction).
Mechanical resonance frequencies of this system are 1.5 Hz in the vertical
and 2.0/2.5 Hz in the horizontal directions, respectively.
Four support pillars of the table are filled with sand
for vibration damping.
Vibrational and acoustic noises generated by vacuum pumps for DR operation
are eliminated by installing them on a rigid frame with rubber springs
and dampers in an acoustic shielded box.
Soft bellows are inserted in pumping lines near the top of the cryostat.
Especially, T-shaped bellows are used for the still pumping line
(76 mm in diameter) of DR.
To increase the rigidity of the DR unit in the IVC,
we inserted additional thermal isolation supports between different
temperature stages below the 1 K pot.
%
%
%%%%%%% Experimental Results %%%%%%%%%%%%%%%%%%%%%
\section{TEST RESULTS}

\subsection{Sample/tip exchange}

\begin{figure}[h]
\includegraphics[width=0.85\linewidth]{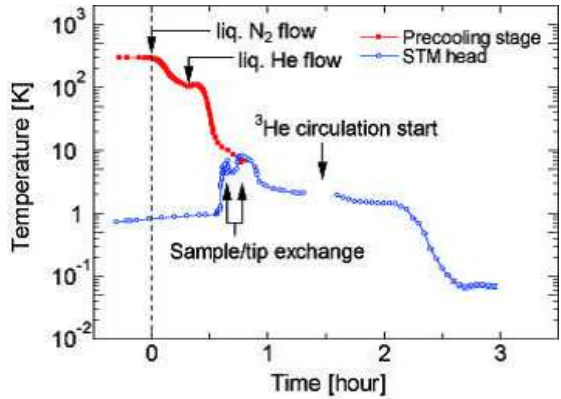}
\caption{\label{tipchange}
Time evolutions of temperatures of the precooling stage and STM head
during the sample/tip exchange procedure.}
\end{figure}

Figure \ref{tipchange} shows time evolutions of temperatures of the
precooling stage and STM head during the sample/tip exchange procedure.
Before inserting the sample/tip, we first stop the $^3$He circulation
of DR.
Although the STM head is warmed temporarily up to about 10 K by
the insertion, it quickly cools back to 2 K due to a large heat
capacity of $^3$He--$^4$He mixture liquid in the DR unit.
By restarting the $^3$He circulation, it cools back below 70 mK
in only 1.5 hours.
It takes about 15 hours more to reach the lowest temperature below 30 mK.
The quick turnaround back to near the base temperature within 3 hours
in total is one of extraordinary features of this ULT-STM.

\subsection{Basic performance of the ULT-STM}
To demonstrate the basic performance of ULT-STM,
we carried out test measurements using a 2H-NbSe$_2$ sample.
2H-NbSe$_2$ is a well-studied material in which the superconducting
phase ($T_{\rm c}$=7.2 K) and the charge density wave (CDW) phase
($T_{\rm CDW}$=33 K) coexist. We used electrochemically
etched W-wires as STM tips for these measurements.

\begin{figure}[h]
\includegraphics[width=1.0\linewidth]{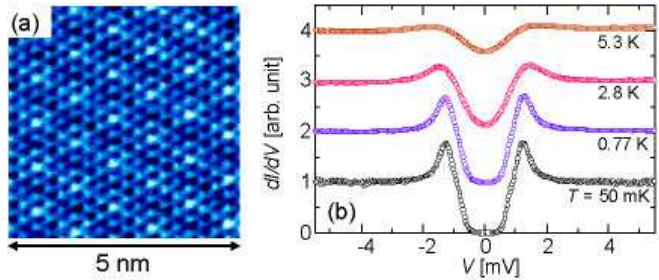}
\caption{\label{nbse2topodidv}
(a) Constant current image of the surface Se atoms of a cleaved
2H-NbSe$_2$ sample obtained at $T=170$ mK
($5\times5$ nm$^2$, $V=-10$ mV, $I=0.20$ nA).
A clear atomic corrugation and a CDW modulation are observed.
(b) Tunnel spectra obtained at various temperatures
($V=-6.0$ mV, $I=0.20$ nA).
The $dI/dV$ curves are offset by 1 unit steps for clarity.}
\end{figure}

Figure \ref{nbse2topodidv}(a) shows a constant current image of
a cleaved surface of 2H-NbSe$_2$ at $T=170$ mK.
The atomic corrugation of surface Se as well as
the CDW modulation are clearly visible.
In the spectroscopy mode, we measured a set of tunnel spectra
averaged over a $2\times2$ nm$^2$ area at various temperatures
(Fig.~\ref{nbse2topodidv}(b)).
An expected temperature dependence of the superconducting energy gap
is observed here.
The modulation frequency used here is 511.7 Hz, and
the modulation amplitude is 20 and 50 $\mu$V$_{\rm rms}$ for
$T\le0.77$ K and $T\ge2.8$ K, respectively.

\begin{figure}[h]
\includegraphics[width=1.0\linewidth]{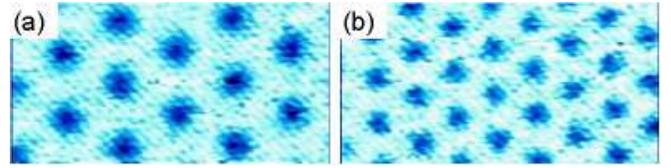}
\caption{\label{nbse2vortex}
$dI/dV$ images of 2H-NbSe$_2$ in magnetic fields of (a) 0.2 T and
(b) 0.4 T showing the Abrikosov (triangular) lattices of the quantized
vortices ($T=100$ mK, $500\times250$ nm$^2$, $V=-1.0$ mV, $I=40$ pA).
The dark spots represent the vortex cores.}
\end{figure}

In Figs.~\ref{nbse2vortex}(a)(b), we show $dI/dV$ images
in the presence of magnetic fields of 0.2 and 0.4 T
on the same sample surface as in Fig.~\ref{nbse2topodidv}.
As the bias voltage is tuned at one of the coherence peaks of the
gap structure for these images, the dark spots
represent the vortex cores \cite{hess2}.
Note that the scan area ($500\times250$ nm$^2$) here is much larger than
that ($5\times5$ nm$^2$) in Fig.~\ref{nbse2topodidv}.
The Abrikosov (triangular) lattices of the quantized vortices
are formed without distortions.

To evaluate the energy resolution $\delta E$ in the tunnel spectroscopy
(TS) mode quantitatively, we performed TS measurements of conventional
superconductors such as Ta, Nb, $etc$. where the standard BCS theory
is applicable.
The normalized tunnel conductance ($G_{\rm NS}$) of a normal
metal--superconductor (N--S) junction at finite temperatures
is given by
\cite{tinkham}:
\begin{equation}
G_{\rm NS}  =  \int_{-\infty}^{\infty}\frac{N_{\rm s}(E)}{N(0)}
\left[-\frac{\partial f(E+eV,T)}{\partial (eV)} \right]dE,
\label{eq:bcs}
\end{equation}
where,
\begin{equation}
\frac{N_{\rm s}(E)}{N(0)}  =  \left\{
\begin{array}{ll}
\frac{|E|}{\sqrt{E^2-\Delta(T)^2}} & (|E|>\Delta(T)) \\
0 & (|E|<\Delta(T)).
\end{array}
\right.
\end{equation}
Here, $N_{\rm s}(E)$ is the density of states of superconductor,
and $N(0)$ is that of normal metal at the Fermi energy $E_F$.
$E$ is the energy measured from $E_F$, and
$f(E,T)$ is the Fermi distribution function at finite $T$.
$\Delta(T)$ is the $T$-dependent superconducting energy gap.

\begin{figure}[h]
\includegraphics[width=0.8\linewidth]{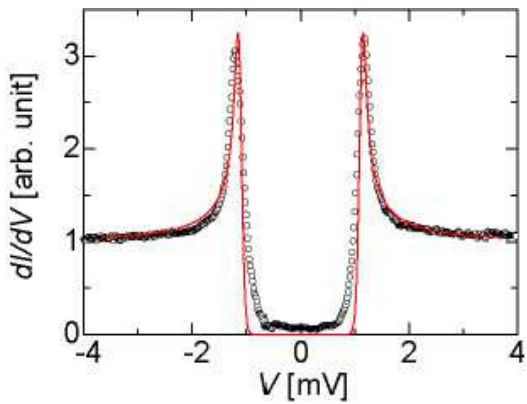}
\caption{\label{nb}
Open circles: tunnel spectrum obtained with a Nb tip on a Au film
($T=52$ mK, $V=-6.0$ mV, $I=2.0$ nA,
$V_{\rm mod}=20$ $\mu$V$_{\rm rms}$, $f_{\rm mod}=511.7$ Hz).
Solid line: calculated conductance using Eq.~(\ref{eq:bcs})
with $\Delta(0)=1.1$ meV and $T_{\rm eff}=350$ mK (see text).
The fitted line deviates from the data substantially
at energies near $\pm1$ meV inside the superconducting gap
as was observed by the previous workers \cite{hamburg,pan2}.
The deviation seems to be sensitively influenced by tip
conditions such as the composition of tip apex, {\it etc.}}
\end{figure}

Figure \ref{nb} shows a measured tunnel spectrum at $T=52$ mK with
a mechanically sharpened superconducting Nb tip (0.3 mm in diameter) on
a normal metallic Au film (120 nm thick) evaporated on a mica substrate.
The solid line is a calculated conductance using Eq.~(\ref{eq:bcs})
with $\Delta(0)=1.1$ meV and $T=350$ mK.
This $T$ value is much higher than the temperature indicated
by the thermometer, since this represents not only thermal broadening
but also smearing due to electrical noises $\delta V$ in measuring circuits.
Thus we call this effective temperature $T_{\rm eff}$ to distinguish
from actual temperature of the sample and tip.
The $T_{\rm eff}$ ($=350$ mK) can be converted to $\delta E$
($\simeq 100$ $\mu$eV) using the relation
$\delta E \simeq 3.5k_{\rm B}T_{\rm eff}$ \cite{wolf}.

We estimated $\delta V \approx 100$ $\mu$V independently by measuring a
superconductor--superconductor (S--S) tunnel spectrum of Ta--Ta.
Note that $\delta V$ is a dominant factor of $\delta E$ in the S--S
tunnel spectra at temperatures well below $T_{\rm c}$ \cite{rodrigo},
{\it i.e.}, $\delta E \simeq e\delta V$ ($\approx 100$ $\mu$eV).
Thus we conclude that the energy resolution in our TS measurements
is almost determined by the electrical noises not, for instance,
by the hot electron effect.

\subsection{Stability against magnetic field sweep}
\begin{figure}[h]
\includegraphics[width=0.9\linewidth]{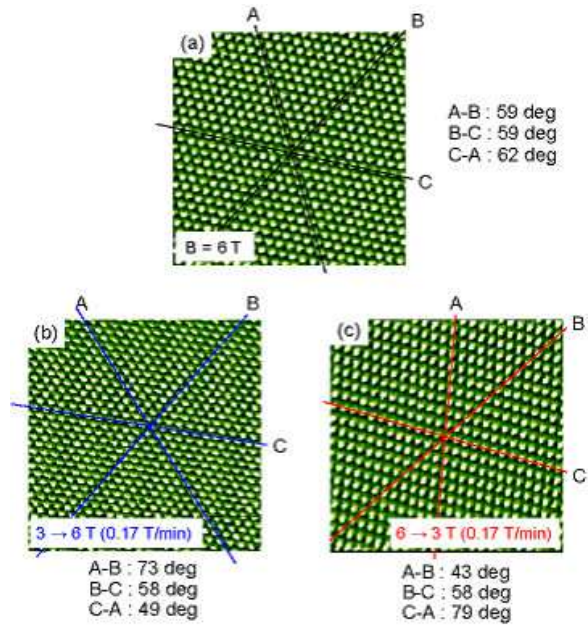}
\caption{\label{hopgsweep}
STM images of an HOPG surface obtained (a) in a stationary field of 6 T,
(b) during field sweeps from 3 to 6 T and (c) from 6 to 3 T
with a speed of 0.17 T/min.
The three lines on each image denote symmetry axes of triangular lattice
($T=1.7$ K, $5\times5$ nm$^2$, $V=200$ mV, $I=0.24$ nA).
}
\end{figure}

Figure \ref{hopgsweep}(a) is an STM image of a surface of highly
oriented pyrolytic graphite (HOPG) obtained at $T=1.7$ K in $B=6$ T.
The image shows a good atomic resolution without any distortions,
and this is always true at any stationary fields from 0 to 6 T.
What is unusual for this ULT-STM is that similarly clear images are
obtained even during sweeping magnetic field from 3 to 6 T and vice versa
(Fig.~\ref{hopgsweep}(b)(c)).
These demonstrate that the system is surprisingly insensitive to the
application of magnetic field, which is another distinguished
performance of this system.
Slight deformations ($< \pm 20^{\circ}$) of the atomic rows are,
however, noticed in the images obtained during the field sweep.
This is presumably due to a weak magnetic force produced by a tiny
magnetization of the STM head.

\subsection{Sample preparation in UHV}
\begin{figure}[h]
\includegraphics[width=0.75\linewidth]{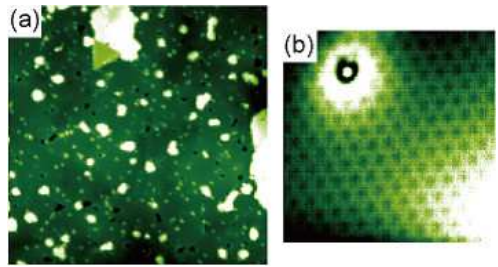}
\caption{\label{inas}
STM images of an InAs(111) surface after removing the As passivation layer
in the manner described in the text.
The images are obtained after cooling to 30 mK.
(a) The triangle in the upper middle is a plane of a stacking fault
tetrahedron ($200\times200$ nm$^2$, $V=0.50$ V, $I=0.20$ nA).
Bright spots are probably corresponding to an As residue.
(b) An STM image of the $(2 \times 2)$ surface reconstruction
($10\times10$ nm$^2$, $V=0.50$ V, $I=0.40$ nA).}
\end{figure}

Taking advantage of the UHV compatibility of this ULT-STM, we made
STM imaging and STS measurements on quasi zero-dimensional (0D)
and two-dimensional (2D) electron systems in an InAs film.
The InAs film (30--100 nm thick) was grown on a GaAs(111)A
substrate by the same molecular-beam epitaxy method as that
described by Kanisawa {\it et al.} \cite{kanisawa} in a separate UHV
chamber. The InAs surface was passivated {\it in situ}
by evaporation of an As film.
Then the sample was transferred to the ULT-STM through an temporary
exposure to air.
We succeeded in retaining the clean surface after removing the As passivation
layer by heating up to 380 $^\circ$C in UHV.
Figure \ref{inas} shows STM images of an InAs surface obtained in
such a way.
A stacking-fault tetrahedron, in which the quasi 0D electron system
is confined, (Fig.~\ref{inas}(a)) and individual surface In atoms
with the $(2 \times 2)$ reconstructed pattern are clearly
visible (Fig.~\ref{inas}(b)).
In addition, the Landau quantization of the 2D electron system is
observed in TS measurements at temperatures below 30 mK
in magnetic fields up to 6 T \cite{niimiinas}.
These measurements became possible only using the present ULT-STM
with the full UHV compatibility.

%
%
%%%%%% other thing %%%%%%%%%%%%%%%%
%\section{other thing}
%
%
%
%%%%%%%%%%%%%%% Summary %%%%%%%%%%%%%%%%%
%\section{SUMMARY}
%
%
%%%%%%%%%%%% Acknowledgments %%%%%%%%%%%%%%%%%%
\section*{Acknowledgments}
This work was financially supported by Grant-in-Aid for Specially
Promoted Research (No.~10102003) and Grant-in-Aid for Scientific
Research on Priority Areas (Nos.~17071002,~17071007) from MEXT, Japan,
Exploratory Research for Advanced Technology Project of JST
(Tarucha Mesoscopic Correlation Field Project) and
Grant-in-Aid for Young Scientists (A) from JSPS (No.~17684011).
We are grateful to Y. Hirayama and K. Kanisawa for supplying us
the InAs thin layer samples.
We thank I. Ueda and T. Shishido for their technical assistance
in the early stage of this work and Unisoku Co., Ltd. for their
technical support in designing and assembling the STM head.
Encouraging discussions with S. Tarucha and Y. Iye throughout this work
were helpful for us.
T.M. and Y.N. acknowledge the JSPS Research program for Young Scientists.
%
%
%
%%%%%%%%%%%%%%%%%%%%%%%%%%%%%%%%%%%%%%%%
% now the references. delete or change fake bibitem. delete next three
%   lines and directly read in your .bbl file if you use bibtex.

% Put the label that you will use with \ref{} command
% in the braces of the \label{} command.
%
%

\begin{thebibliography}{30}
\bibitem{oxford} Model Kelvinox-100, Oxford Instruments.
The cooling power is 100 $\mu$W at 100 mK, and the base temperature
without heat load is 9.4 mK.
\bibitem{unisokustm} The STM head was designed and assembled in
collaboration with Unisoku Co., Ltd.
\end{thebibliography}
\end{document}